\documentclass[twocolumn,showpacs,amsmath,amssymb,nofootinbib,amsfonts,showkeys]{revtex4}
\usepackage[toc,page]{appendix}
\usepackage{graphicx}
\usepackage{caption}
\usepackage{subfig}
\usepackage{bm}
\usepackage{mathtext}
\usepackage[T2A]{fontenc}
\usepackage[utf8]{inputenc}
\usepackage[english,ukrainian]{babel}

\begin{document}

\title{Evolution of cosmological perturbations \\
in the models with interacting dynamical dark energy}
\author{R. Neomenko}
 \email{oz.rik@hotmail.com}
\author{B. Novosyadlyj}
\affiliation{Astronomical Observatory of Ivan Franko National University of Lviv, Kyryla i Mefodiya Street, 8, Lviv, 79005, Ukraine}

\date{\today}

\begin{abstract}

Evolution of cosmological perturbations is considered in the model with dynamical dark energy which non-gravitationally interacts with dark matter. The dark energy equation of state parameter (EoS) is varying in time and is parameterized by its adiabatic sound speed. Such model of interacting dark energy has advantages over model with constant EoS, because it avoids non-adiabatic instabilities of dark energy at radiation dominated epoch for certain types of interaction in dark sector. The stability conditions for solutions of equations of dark energy perturbations were derived. The impact of strength of additional interaction between dark components on the evolution of density and velocity perturbations in them is analysed for quintessence and phantom types of dark energy.
\end{abstract}



\maketitle

\section*{INTRODUCTION}
All current observational cosmology data indicate that beside visible matter (which consists of the particles of Standard Model) there is dark sector of unknown components - dark energy \cite{Riess1998, Perlmutter1999} and dark matter. The cosmological model which more or less fits the observational data is $\Lambda$CDM model. However this model has the problems of interpretations of the observational data appeared in the post Planck cosmology \cite{Colin2019, Kang2020, DiValentino2020, Nadathur2020}. So we can make our model more complicated, where there may exist some non-gravitational coupling between dark components which make significant impact on expansion dynamics of Universe and formation of its large-scale structure. Such cosmological models were studied in works \cite{Amendola2000, Zimdahl2001, Amendola2007, Bolotin2013, Chimento2010}. As it follows from \cite{DiValentino2017, Rui2018, DiValentino2019} they can resolve some problems.

It is known that in the dark energy models with constant equation of state parameter (EoS) the instabilities of cosmological perturbations appear at the radiation-dominated epoch if additional non-gravitational interaction is present \cite{Valiviita2008, Jackson2009}. The solution of this problem as mentioned in the papers \cite{Valiviita2008} and \cite{Majerotto2010} can be the dynamical dark energy with varying EoS \cite{Sharov2017, WYang2018, Yang2019a, Bonici2019, Goswami2019}.

In this paper we study the evolution of cosmological perturbations in a three component Universe which consists of dynamical dark energy, dark matter and primordial electromagnetic radiation. The dark components non-gravitationally interact with each other (DE-DM interaction). The dark energy is represented by a model with varying EoS parameter $w$ which is parameterized by adiabatic sound speed $c_{a}$ and EoS parameter at present time $w_{0}$ \cite{Novosyadlyj2010, Novosyadlyj2012, Sergijenko2015}. The dark matter is described by a pressureless ideal fluid model; however effective pressure as a result of the energy-momentum exchange between the dark energy and the dark matter can arise.

The DE-DM interaction term in general-covariant conservation equations for dark energy and dark matter is proportional to some function of the energy densities of dark components. In this paper, we explore the cosmological models with the simplest interactions, which are linearly dependent on the density of dark components.

\section{Model of interacting dynamical dark energy}

We consider the spatialy flat homogeneous and isotropic Universe with Friedman-Lema\^{\i}tre-Robertson-Walker (FLRW) metric:
\begin{eqnarray}\label{ds}
 ds^{2}=g_{ik}dx^idx^k=a^{2}(\eta)[d\eta^{2}-\delta_{\alpha\beta}dx^{\alpha}dx^{\beta}],
\end{eqnarray}
where $g_{ik}$ is metric tensor, $a(\eta)$ is scale factor, $\eta$ is conformal time, which related to physical time $t$ by $dt=a(\eta)d\eta$. Hereafter we assume the speed of light $c$ equals unity. At present time $a(\eta_0)=1$. Each of the components - dark energy, dark matter and black-body electromagnetic radiation, is described by an ideal fluid approximation with energy-momentum tensor:
\begin{equation}\label{Tik}
 T_{(N)i}^{k}=(\rho_{(N)}+p_{(N)})u_{(N)i}u_{(N)}^{k}-p_{(N)}\delta_{i}^{k},
\end{equation}
where $\rho_{(N)}$ is energy density of $N$ component, $p_{(N)}$ is its pressure, $u_{(N)i}$ is 4-vector of velocity. The equation of state of each component is given by $p_{(N)}=w_{(N)}\rho_{(N)}$, where for dark energy $w_{de}=w$, for dark matter $w_{dm}=0$ and for radiation $w_{r}=1/3$.

The general-covariant conservation law $\sum_{N}T_{(N)i}^{k}=0$ gives the following equations for the evolution of the energy and momentum densities of the dark components with DE-DM interaction
\begin{eqnarray}
 & & T_{(de)i;k}^{k}=J_{(de)i}, \label{tq1} \\
 & & T_{(dm)i;k}^{k}=J_{(dm)i}, \label{tq2}
\end{eqnarray}
where $J_{i(de,dm)}$ describes the DE-DM interaction between the components, the semicolon denotes the covariant derivative. It follows from the energy-momentum conservation law that $J_{(de)i}+J_{(dm)i}=0$, so we can just put $J_{(de)i}=-J_{(dm)i}=J_{i}$. For the unperturbed medium with metric (\ref{ds}) the conservation equations are as follows:
\begin{eqnarray}
  & & \dot{\bar{\rho}}_{de}+3aH(1+w)\bar{\rho}_{de}=\bar{J}_{0}, \label{eq de} \\
  & & \dot{\bar{\rho}}_{dm}+3aH\bar{\rho}_{dm}=-\bar{J}_{0}, \label{eq dm}
\end{eqnarray}
where the dot denotes the derivative with respect to conformal time $\eta$, $H=\dot{a}/a^{2}$ is the Hubble parameter which characterizes the expansion rate of the Universe, $\bar{J}_{0}$ is the background 0-component of $J_{i}$.

Also from Einstein's equations
\begin{equation}\label{GT}
 R_{ik}-\frac{1}{2}Rg_{ik}=8\pi G\sum_{N}T_{(N)ik}
\end{equation}
one can obtain equations for the expansion dynamics of the Universe in FLRW metric (\ref{ds}):
\begin{eqnarray}
  & & H^{2}=\frac{8\pi G}{3}\sum_{N}\bar{\rho}_{(N)}, \label{H2} \\
  & & qH^{2}=\frac{4\pi G}{3}\sum_{N}(\bar{\rho}_{(N)}+3\bar{p}_{(N)}), \label{qH2}
\end{eqnarray}
where $q\equiv-\frac{\ddot{a}}{a^{3}H^{2}}+1$ is the deceleration parameter.

The dark energy with the coupling linearly dependent on the density of dark matter is unstable when $w$ is close to $-1$ at the scales much larger then the Hubble horizon in the radiation dominated epoch \cite{Valiviita2008}. To avoid this, we consider a more general model of dynamical dark energy, where $w$ is variable in time like in paper \cite{Majerotto2010}. In this work, we consider the dynamical dark energy with adiabatic sound speed $c_{a}^{2}\equiv\dot{\bar{p}}_{de}/\dot{\bar{\rho}}_{de}=const$ \cite{Novosyadlyj2010, Novosyadlyj2012, Sergijenko2015}. Such parameterization  gives us a possibility to explore larger numbers of the valuable interacting quintessence and phantom dark energy models. This condition and conservation equation (\ref{eq de}) lead to equation:
\begin{equation}\label{dw}
 \frac{dw}{da}=\frac{3}{a}(1+w)(w-c_{a}^{2})-\frac{\bar{J}_{0}}{\bar{\rho}_{de}a^{2}H}(w-c_{a}^{2}).
\end{equation}
For the EoS parameter, we have also the connection to the density of dark energy for an arbitrary form of interaction:
\begin{equation}\nonumber
 w=c_{a}^{2}+\bar{\rho}_{de}^{(0)}\frac{w_{0}-c_{a}^{2}}{\bar{\rho}_{de}}.
\end{equation}
where $w_{0}$ is the EoS parameter in the present epoch. To describe the expansion dynamics of the Universe, we must solve the system of equations (\ref{eq de}), (\ref{eq dm}), (\ref{dw}) and (\ref{H2}).

Also, we must determine in which form the background interaction term $\bar{J}_{0}$ is to be set. Here we assume that in the general case it depends on $H$, $\bar{\rho}_{de}$ and $\bar{\rho}_{dm}$.

For convenience we will rewrite the conservation equations in such form:
\begin{eqnarray}
  & & \dot{\bar{\rho}}_{de}+3aH(1+w+\Pi_{de})\bar{\rho}_{de}=0, \nonumber \\
  & & \dot{\bar{\rho}}_{dm}+3aH(1-\Pi_{dm})\bar{\rho}_{dm}=0, \nonumber
\end{eqnarray}
where
$$\Pi_{de}=-\bar{J}_{0}/(3aH\bar{\rho}_{de}), \quad \Pi_{dm}=-\bar{J}_{0}/(3aH\bar{\rho}_{dm})$$ 
are effective corrections to the EoS parameters of the dark components, which appear as a result of the non-gravitational interaction between the components.

\section{Cosmological perturbation equations for non-minimally coupled dark energy model}

\subsection*{Perturbations of an energy-momentum tensor}

Let us consider the perturbed parts of conservation equations (\ref{tq1}), (\ref{tq2}) and Einstein's equations (\ref{GT}). In the conformal-Newtonian gauge, the perturbed metric is:
\begin{equation}\label{dspsi}
 ds^{2}=a^{2}[(1+2\Psi)d\eta^{2}-(1-2\Psi)\delta_{\alpha\beta}dx^{\alpha}dx^{\beta}].
\end{equation}
The perturbed part of energy-momentum tensor (\ref{Tik}) for each component
\begin{equation}\nonumber
 T_{i}^{k}=\bar{T}_{i}^{k}+\delta T_{i}^{k},
\end{equation}
can be represented by the perturbed density, pressure and 4-velocity:
\begin{eqnarray}
 & & \rho = \bar{\rho}(1+\delta), \quad p=\bar{p}+\delta p, \nonumber \\
 & & u^{i}=\bar{u}^{i}+\delta u^{i}, \nonumber \\
 & & \delta u^{i}=\left(-\frac{\Psi}{a}, \frac{v^{\alpha}}{a}\right), \nonumber
\end{eqnarray}
where $v^{\alpha}\equiv\frac{dx^{\alpha}}{d\eta}$ and $\bar{u}^{i}=(a^{-1}, 0, 0, 0)$. Hence, the components of the perturbed energy-momentum tensor of the perfect fluid are as follows:
\begin{eqnarray}
 & & \delta T_{0}^{0}=\bar{\rho}\delta, \quad \delta T_{0}^{\alpha}=(\bar{\rho}+\bar{p})v^{\alpha}, \nonumber \\
 & & \delta T_{\alpha}^{0}=-(\bar{\rho}+\bar{p})v^{\alpha}, \quad \delta T_{\alpha}^{\beta}=-\delta_{\alpha\beta}\delta p. \nonumber
\end{eqnarray}
The perturbation of the pressure of the dark energy in the conformal-Newtonian gauge can be presented as a sum of the adiabatic and non-adiabatic parts:
\begin{equation}\nonumber
 \delta p_{de}=c_{a}^{2}\bar{\rho}_{de}\delta_{de}+\delta p_{n-ad},
\end{equation}
In the rest frame of dark energy, the perturbation of pressure is:
\begin{equation}\nonumber
 \delta p_{de}^{(rf)}=c_{s}^{2}\bar{\rho}_{de}\delta_{de}^{(rf)},
\end{equation}
where $c_{s}^{2}$ is the effective sound speed of dark energy in its rest frame. Using linear transformations between gauges, one can obtain a general expression for the perturbation of pressure in the conformal-Newtonian gauge:
\begin{eqnarray}
 & & \delta p_{de}=c_{s}^{2}\bar{\rho}_{de}\delta_{de}- \nonumber\\
 & & -(c_{s}^{2}-c_{a}^{2})[3aH(1+w)\bar{\rho}_{de}-J_{0}]\int\vec{v}_{de}d\vec{x}. \nonumber
\end{eqnarray}

\subsection*{Covariant form of DE-DM interaction term}

In \cite{Neomenko2016} we studied the expansion dynamics of the Universe with the DE-DM interactions of three types:
\begin{eqnarray}
 & & \bar{J}_{0}=-3\alpha a\Gamma\bar{\rho}_{cr}, \label{J0cr} \\
 & & \bar{J}_{0}=-3\beta a\Gamma\bar{\rho}_{de}, \label{J0de} \\
 & & \bar{J}_{0}=-3\gamma a\Gamma\bar{\rho}_{dm}, \label{J0dm}
\end{eqnarray}
where $\alpha$, $\beta$ and $\gamma$ are coupling constants, $\rho_{cr}\equiv\frac{3H_{0}^{2}}{8\pi G}$ is critical density. There are two most common choices for $\Gamma$ in the papers devoted to non-minimally coupled dark energy: $\Gamma=const$ \cite{Valiviita2008, SWang2008, Caldera2009, Feng2020} and $\Gamma=H$ \cite{Amendola2007, Bolotin2013, Chimento2010, Jackson2009, Zhou2009, delCampo2006, delCampo2009, Feng2019, Wang2007, Rosenfeld2007}. We analyse the case when $\Gamma=H$.

Here the DE-DM interactions are written for the unperturbed 0-component of $J_{i}$.  In the case with perturbations, it must be written in a more general form for dark components with energy-momentum tensor (\ref{Tik}). Also, it must satisfy the general covariance principle, since the physics of such interaction must not be depended on the change of the reference frame. In this work, we use the covariant form of the DE-DM interaction from \cite{Gavela2010}:
\begin{eqnarray}
 & & J_{i}=-f(\rho_{de}, \rho_{dm})u^{k}_{(T);k}u_{(dm)i}. \label{JTdm}
\end{eqnarray}
Here the energy density of each dark component is defined in the reference frame of dark matter:
\begin{equation}\nonumber
 \rho_{(M)}=T_{(M)ik}u_{(dm)}^{i}u_{(dm)}^{k},
\end{equation}
where index $M=(de, dm)$. The background part of this interaction has the form (\ref{J0cr})-(\ref{J0dm}). As a result interaction $J_{i}$ in the conformal-Newtonian gauge is:
\begin{eqnarray}
 & & J_{0}=-3aH\bar{f}(\bar{\rho}_{de}, \bar{\rho}_{dm})(1+\varepsilon)+ \nonumber \\
 & & +\bar{f}(\bar{\rho}_{de}, \bar{\rho}_{dm})\left(3\dot{\Psi}-\sum_{\alpha}\frac{\partial v_{T}^{\alpha}}{\partial x^{\alpha}}\right), \label{JP0dm} \\
 & & J_{\alpha}=3aH\bar{f}(\bar{\rho}_{de}, \bar{\rho}_{dm})v_{dm}^{\alpha} ,\label{JPadm}
\end{eqnarray}
where $\varepsilon\equiv \delta f/\bar{f}$ and $v_{T}^{\alpha}=\sum_{N}(\bar{\rho}_{(N)}+\bar{p}_{(N)})v_{(N)}^{\alpha}/\sum_{N}(\bar{\rho}_{(N)}+\bar{p}_{(N)})$. If the coupling constant goes to zero then the energy-momentum transfer between dark components vanishes. If perturbations vanish then (\ref{JP0dm})-(\ref{JPadm}) reduce to the background form of (\ref{J0cr})-(\ref{J0dm}). In this paper, we study such partial cases of interactions (\ref{JP0dm}), (\ref{JPadm}) with functions $\bar{f}$, $\varepsilon$:
\begin{eqnarray}
 & & \bar{f}=\alpha\bar{\rho}_{cr}, \quad \varepsilon=0, \label{fcr} \\
 & & \bar{f}=\beta\bar{\rho}_{de}, \quad \varepsilon=\delta_{de}, \label{fde} \\
 & & \bar{f}=\gamma\bar{\rho}_{dm}, \quad \varepsilon=\delta_{dm}, \label{fdm}
\end{eqnarray}
Analytical solutions of background equations (\ref{eq de}), (\ref{eq dm}) and (\ref{dw}) were obtained for the DE-DM interactions with the linear dependence of $\bar{f}$ on the densities of dark components and studied in details in the papers \cite{Neomenko2016, Neomenko2017}.

\subsection*{Cosmological perturbation equations}

Conservation equations (\ref{tq1}), (\ref{tq2}) and Einstein's equation (\ref{GT}) together with the DE-DM interaction (\ref{JP0dm}), (\ref{JPadm}) give the system of equations for the evolution of the density and velocity perturbations of dark matter, dark energy and radiation as well as the metric perturbation function $\Psi$
\begin{widetext}
\begin{eqnarray}
 & & \dot{\delta}_{de}=-3aH(c_{s}^{2}-w)\delta_{de}+3(1+w)\dot{\Psi}+(1+w)[k^{2}+9a^{2}H^{2}(c_{s}^{2}-c_{a}^{2})]V_{de}+ \nonumber \\
 & & +\Pi_{de}[3aH(\delta_{de}-\varepsilon)+3\dot{\Psi}+k^{2}V_{T}+9a^{2}H^{2}(c_{s}^{2}-c_{a}^{2})V_{de}], \label{pedde}\\
 & & \dot{V}_{de}=-aH(1-3c_{s}^{2})V_{de}-\frac{c_{s}^{2}}{1+w}\delta_{de}-\Psi+3aH\frac{\Pi_{de}}{1+w}[(1+c_{s}^{2})V_{de}-V_{dm}], \label{pevde}\\
 & & \dot{\delta}_{dm}=3\dot{\Psi}+k^{2}V_{dm}-\Pi_{dm}[3aH(\delta_{dm}-\varepsilon)+3\dot{\Psi}+k^{2}V_{T}], \label{peddm}\\
 & & \dot{V}_{dm}=-aHV_{dm}-\Psi, \label{pevdm}\\
 & & \dot{\delta}_{r}=4\dot{\Psi}+\frac{4}{3}k^{2}V_{r}, \label{pedr}\\
 & & \dot{V}_{r}=-\Psi-\frac{\delta_{r}}{4}, \label{pevr}\\
 & & \dot{\Psi}=-aH\Psi-\frac{3}{2}a^{2}H^{2}(1+w_{T})V_{T}, \label{pepsi}
\end{eqnarray}
\end{widetext}
where $\delta$ is the Fourier amplitude of the density perturbation, $V\equiv -i(\overrightarrow{k},\overrightarrow{v})/k^{2}$ is proportional to the Fourier amplitude of the velocity perturbation, $\Psi$ is the Fourier amplitude of the metric perturbation and $w_{T}=\sum_{N}w_{(N)}\bar{\rho}_{(N)}/\sum_{N}\bar{\rho}_{(N)}$. This system of equations should be integrated with equations (\ref{eq de}), (\ref{eq dm}), (\ref{H2}), containing the background energy densities of each component.The equations (\ref{pedde})-(\ref{pevdm}) in the non-interacting case are the same as the perturbation equations in the linear approximation in paper \cite{NTK}.

\section{Background asymptotic evolution of dark components}

\subsection*{Interaction $\bar{f}=\alpha\bar{\rho}_{cr}$}

Let us consider the expressions for the background energy densities of dark components $\bar{\rho}_{de}$, $\bar{\rho}_{dm}$ \cite{Neomenko2016} in the model of interacting dark energy with a coupling to dark matter independent of its densities (\ref{fcr}):
\begin{eqnarray}
 & & \bar{\rho}_{de}(a)=\bar{\rho}_{de}^{(0)}\frac{(1+w_{0})a^{-3(1+c_{a}^{2})}-w_{0}+c_{a}^{2}}{1+c_{a}^{2}}- \nonumber \\
 & & -\alpha\bar{\rho}_{cr}\frac{1-a^{-3(1+c_{a}^{2})}}{1+c_{a}^{2}}, \nonumber \\
 & & \bar{\rho}_{dm}(a)=\bar{\rho}_{dm}^{(0)}a^{-3}+\alpha\bar{\rho}_{cr}(1-a^{-3}), \nonumber
\end{eqnarray}
The dark energy is quintessential when $c_{a}^{2}>-1$.

In the radiation dominated epoch, we have such asymptotes for the quintessence dark energy: $w=c_{a}^{2}$, $\Pi_{de}=0$, $\Pi_{dm}=0$.

The dark energy of a phantom type when $c_{a}^{2}<-1$. The asymptotes in the early epoch are:
\begin{eqnarray}
 & & w=\frac{\alpha c_{a}^{2}-(w_{0}-c_{a}^{2})\Omega_{de}}{\alpha+(w_{0}-c_{a}^{2})\Omega_{de}}, \quad \Pi_{de}=-\frac{\alpha(1+c_{a}^{2})}{(w_{0}-c_{a}^{2})\Omega_{de}+\alpha}, \nonumber \\
 & & \Pi_{dm}=0.\nonumber
\end{eqnarray}

The condition of the positive energy densities of dark components, $\bar{\rho}_{dm}\ge0$ and $\bar{\rho}_{de}\ge0$, leads to the constraint for the value of interaction parameter \cite{Neomenko2016}.

\subsection*{Interaction $\bar{f}=\beta\bar{\rho}_{de}$}

For the interaction proportional to the density of dark energy $\rho_{de}$ (\ref{fde}) we have such expressions for the energy densities of dark components \cite{Neomenko2016}:
\begin{eqnarray}
 & & \bar{\rho}_{de}(a)=\bar{\rho}_{de}^{(0)}\frac{(1+w_{0}+\beta)a^{-3(1+c_{a}^{2}+\beta)}-w_{0}+c_{a}^{2}}{1+c_{a}^{2}+\beta}, \nonumber \\
 & & \bar{\rho}_{dm}(a)=\bar{\rho}_{dm}^{(0)}a^{-3}+\beta\bar{\rho}_{de}^{(0)}\displaystyle\biggl[\biggl(\frac{A}{c_{a}^{2}+\beta}+B\biggr)a^{-3}- \nonumber \\
 & & -\frac{A}{c_{a}^{2}+\beta}a^{-3(1+c_{a}^{2}+\beta)}-B\biggr], \nonumber \\
 & & A=\frac{1+w_{0}+\beta}{1+c_{a}^{2}+\beta}, \quad B=\frac{w_{0}-c_{a}^{2}}{1+c_{a}^{2}+\beta}. \nonumber
\end{eqnarray}

In this case the dark energy is quintessential, when $\beta>-1-c_{a}^{2}$. At the radiation dominated epoch, for a quintessence model with the conditions of positivity of densities of dark components taken into account we have such asymptotes: $w=c_{a}^{2}$, $\Pi_{de}=\beta$, $\Pi_{dm}=0$.

For phantom model ($\beta<-1-c_{a}^{2}$) we have: $w=-1-\beta$, $\Pi_{de}=\beta$, $\Pi_{dm}=0$.

\subsection*{Interaction $\bar{f}=\gamma\bar{\rho}_{dm}$}

For the interaction proportional to the density of dark matter $\rho_{dm}$ (\ref{fdm}) the expressions for densities $\rho_{de}$, $\rho_{dm}$ are as follows \cite{Neomenko2016}
\begin{eqnarray}
 & & \bar{\rho}_{de}(a)=\bar{\rho}_{de}^{(0)}\bigg[\frac{(1+w_{0})a^{-3(1+c_{a}^{2})}+c_{a}^{2}-w_{0}}{1+c_{a}^{2}}+ \nonumber \\
 & & +\gamma\frac{\Omega_{dm}}{\Omega_{de}}\frac{1-a^{3(c_a^2+\gamma)}}{c_{a}^{2}+\gamma}a^{-3(1+c_{a}^{2})}\bigg]. \nonumber \\
 & & \bar{\rho}_{dm}(a)=\bar{\rho}_{dm}^{(0)}a^{-3(1-\gamma)}, \nonumber
\end{eqnarray}
For the density of dark energy to be positive, the following conditions must be satisfied: $\gamma>0$, $c_{a}^{2}+\gamma<0$. Also, as we see from the expression for $\bar{\rho}_{dm}$, the coupling constant must have small value $\gamma\ll1$ in order to avoid a contradiction with astronomical observations. In the early epoch when $a\rightarrow 0$ the densities of the quintessence and phantom dark energies diverge $\bar{\rho}_{de}\rightarrow\infty$, $\bar{\rho}_{dm}\rightarrow\infty$.

The asymptotes for the quintessence and phantom dark energies in the early epoch are as follows: $w=c_{a}^{2}$, $\Pi_{de}=-c_{a}^{2}-\gamma$, $\Pi_{dm}=\gamma$.

The energy density of radiation, which is in equations (\ref{H2}), (\ref{qH2}) changes over all time in a standard manner $\bar{\rho}_{r}=\bar{\rho}_{r}^{(0)}a^{-4}$.

\begin{figure*}
\captionsetup[subfigure]{labelformat=empty}
\centering
\subfloat[][]{\includegraphics[width=0.45\textwidth]{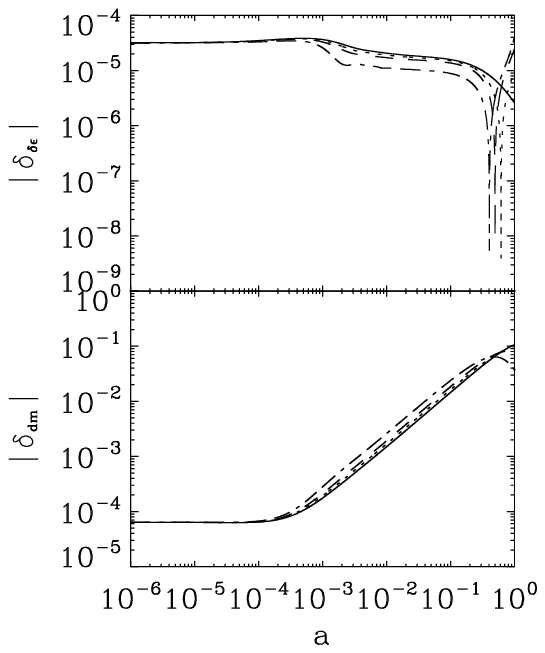}}\quad
\subfloat[][]{\includegraphics[width=0.45\textwidth]{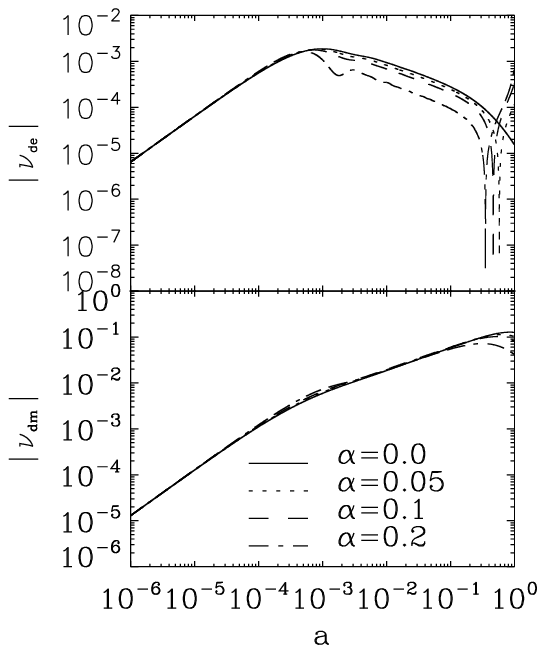}}\\
\subfloat[][]{\includegraphics[width=0.45\textwidth]{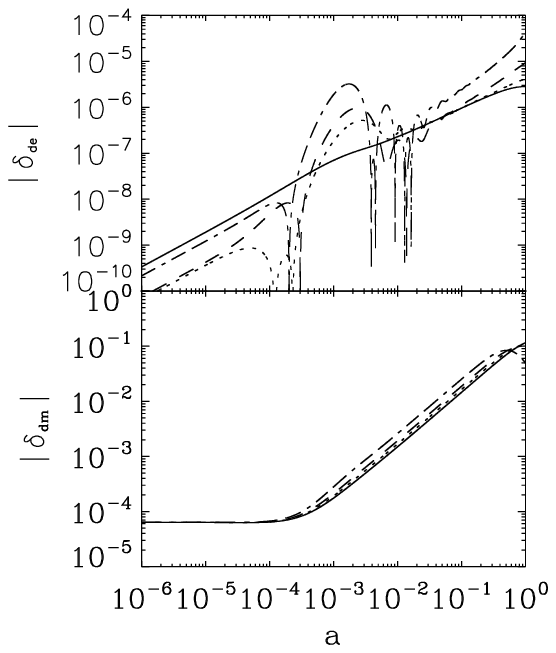}}\quad
\subfloat[][]{\includegraphics[width=0.45\textwidth]{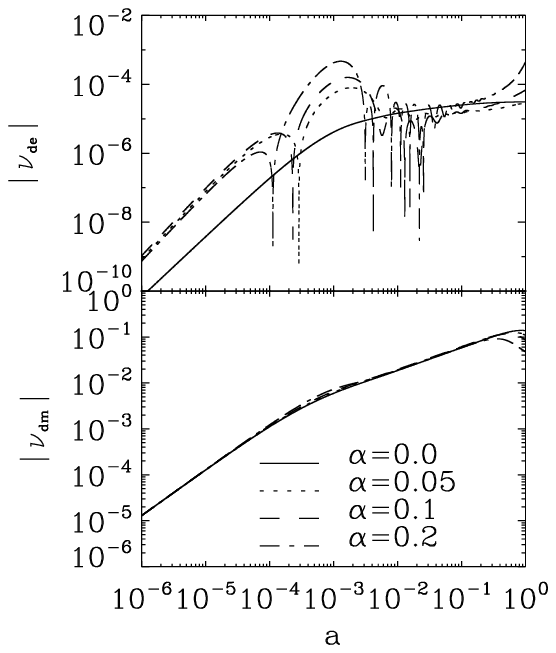}}
\caption{Influence of the DE-DM interaction independent of densities of dark components on the evolution of perturbations of densities and effective momenta of dark energy and dark matter. Here $\Omega_{de}=0.7$, $\Omega_{r}=5.0\cdot10^{-5}$, $k=0.01Mpc$, $c_{s}^{2}=1.0$. For quintessence dark energy (up) $w_{0}=-0.9$, $c_{a}^{2}=-0.5$, for phantom (bottom): $w_{0}=-1.1$, $c_{a}^{2}=-1.25$.}
  \label{fig:cr1}
\end{figure*}

\begin{figure*}
\captionsetup[subfigure]{labelformat=empty}
\centering
\subfloat[][]{\includegraphics[width=0.45\textwidth]{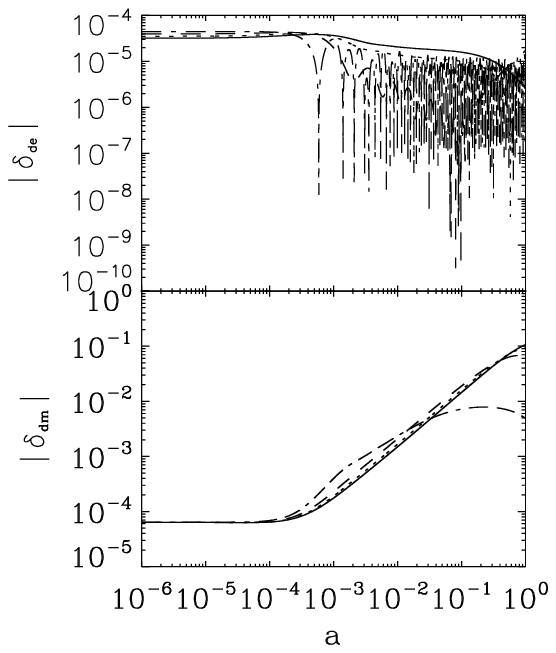}}\quad
\subfloat[][]{\includegraphics[width=0.45\textwidth]{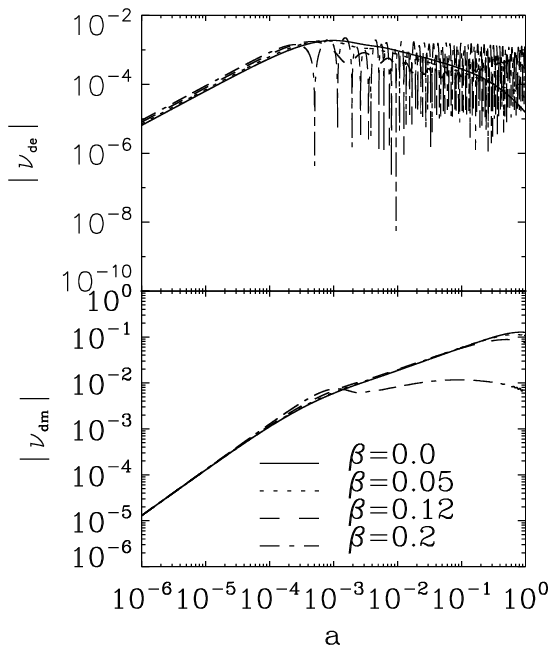}}\\
\subfloat[][]{\includegraphics[width=0.45\textwidth]{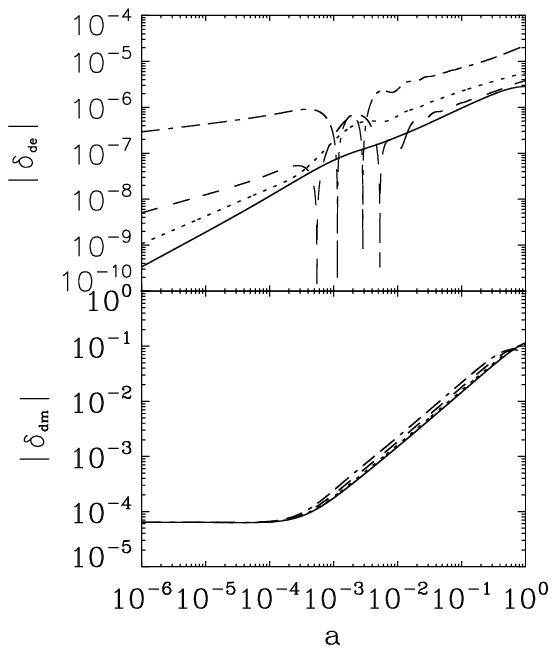}}\quad
\subfloat[][]{\includegraphics[width=0.45\textwidth]{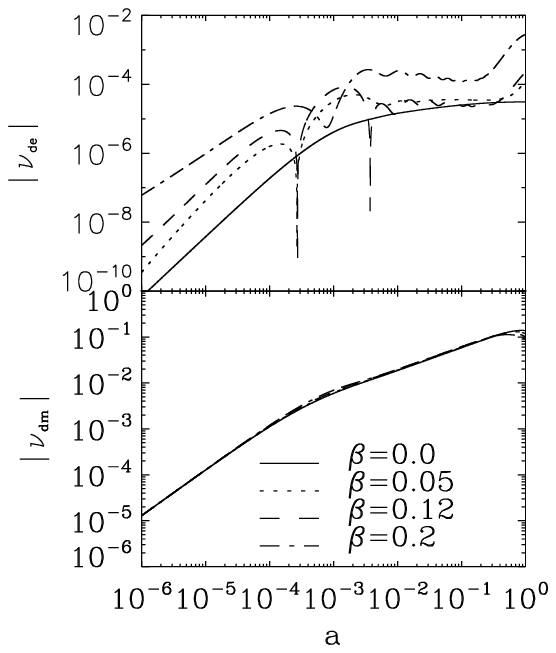}}
\caption{Influence of the DE-DM interaction proportional to $\rho_{de}$ on the evolution of perturbations of densities and effective momenta of dark energy and dark matter. The values of cosmological parameters and parameters of quintessence (up) and phantom (bottom) dark energies are the same as in Fig. \ref{fig:cr1}}
  \label{fig:de1}
\end{figure*}

\begin{figure*}
\captionsetup[subfigure]{labelformat=empty}
\centering
\subfloat[][]{\includegraphics[width=0.45\textwidth]{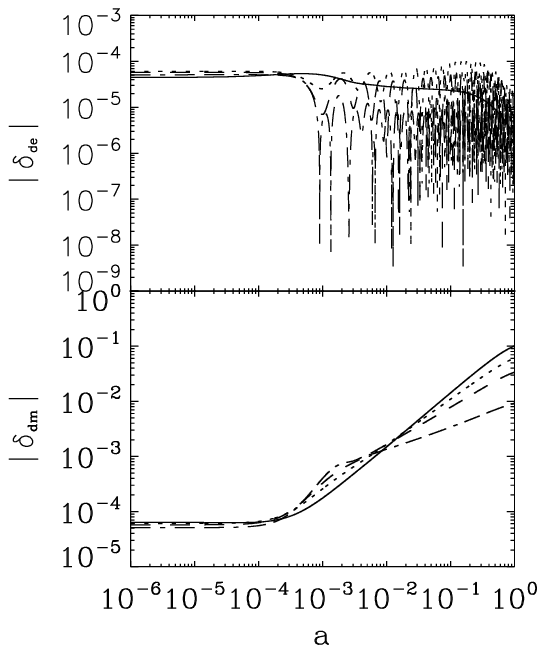}}\quad
\subfloat[][]{\includegraphics[width=0.45\textwidth]{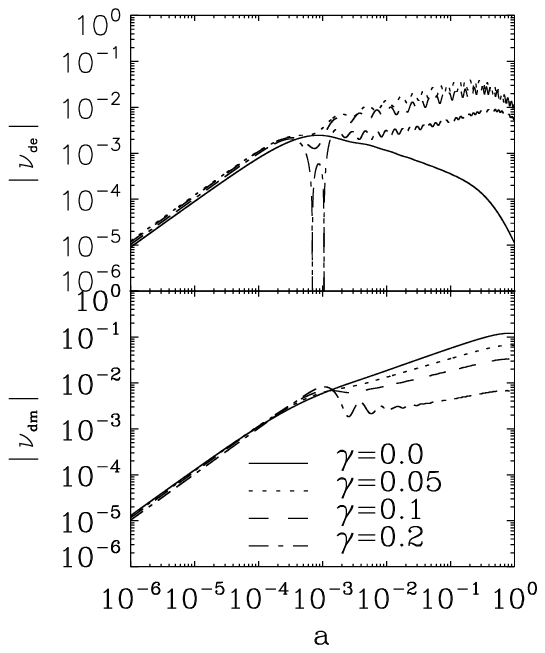}}\\
\subfloat[][]{\includegraphics[width=0.45\textwidth]{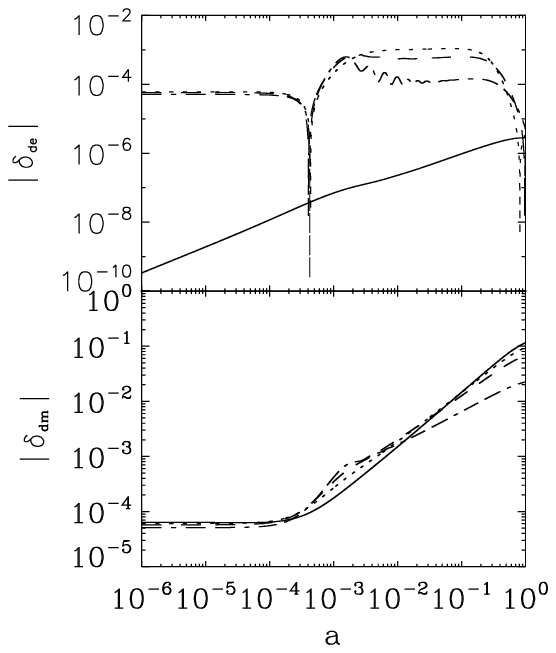}}\quad
\subfloat[][]{\includegraphics[width=0.45\textwidth]{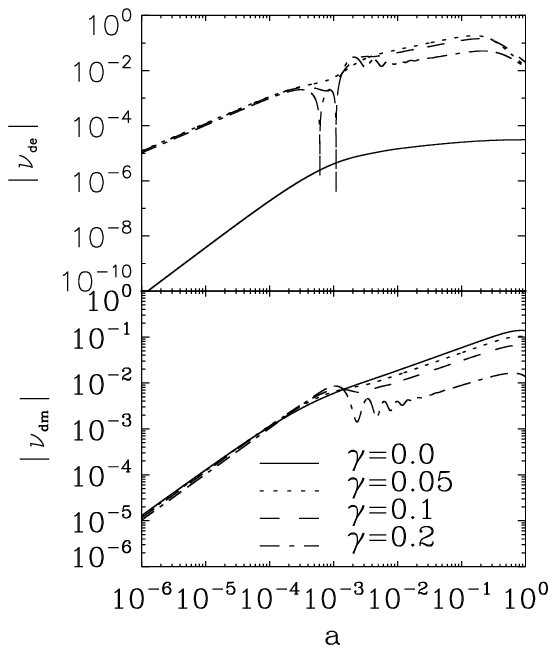}}
\caption{Influence of the DE-DM interaction proportional to $\rho_{dm}$ on the evolution of perturbations of densities and effective momenta of dark energy and dark matter. The values of cosmological parameters and parameters of phantom (bottom) dark energy are the same as in Fig. \ref{fig:cr1}. For quintessence dark energy (up): $w_{0}=-0.9$, $c_{a}^{2}=-0.3$.}
  \label{fig:dm1}
\end{figure*}

\section{Initial and stability conditions for system of perturbation equations}

For scales much larger than the Hubble horizon ($k\eta\ll 1$), in the early epoch the dark energy, dark matter and radiation components behave as an adiabatic fluid. Hence the relative entropy perturbation between arbitrary two components $x$, $y$ is equal to zero:
\begin{equation}\label{sdmr}
 S_{x,y}=aH\left(\frac{\delta_{x}}{(\dot{\bar{\rho}}_{x}/\bar{\rho}_{x})}-\frac{\delta_{y}}{(\dot{\bar{\rho}}_{y}/\bar{\rho}_{y})}\right)=0.
\end{equation}
In the early epoch, the radiation component has a dominant impact on the expansion dynamics of the Universe: $a=H_{0}\sqrt{\Omega_{r}}\eta$ ($\Omega_{r}$ is the relative energy density part of radiation in the present epoch). Moreover, the gravitational potential $\Psi$ of cosmological perturbations is defined by radiation mainly, so, the dark components are dynamically test ones, they practically do not affect it. Hence, at the superhorizon stage when the initial conditions are to be set, equations (\ref{pedde})-(\ref{pepsi}) can be simplified and only the first four equations from the seven can be analyzed because $\Psi=const$ then. Making the transition in equations (\ref{pedde})-(\ref{pevdm}) to the derivative with respect to $N=\ln(k\eta)$, we obtain the perturbed conservation equations in the radiation dominated epoch as follows
\begin{widetext}
\begin{eqnarray}
 & & \frac{d\delta_{de}}{dN}=-3(c_{s}^{2}-w)\delta_{de}+9H_{0}\sqrt{\Omega_{r}}(1+w)(c_{s}^{2}-c_{a}^{2})\tilde{V}_{de}+3\Pi_{de}[\delta_{de}-\varepsilon+3H_{0}\sqrt{\Omega_{r}}(c_{s}^{2}-c_{a}^{2})\tilde{V}_{de}], \label{peNdde} \\
 & & \frac{d\tilde{V}_{de}}{dN}=-(2-3c_{s}^{2})\tilde{V}_{de}-\frac{c_{s}^{2}}{H_{0}\sqrt{\Omega_{r}}(1+w)}\delta_{de}-\frac{1}{H_{0}\sqrt{\Omega_{r}}}\Psi+3\frac{\Pi_{de}}{1+w}[(1+c_{s}^{2})\tilde{V}_{de}-\tilde{V}_{dm}], \label{peNvde} \\
 & & \frac{d\delta_{dm}}{dN}=-3\Pi_{dm}(\delta_{dm}-\varepsilon), \label{peNddm} \\
 & & \frac{d\tilde{V}_{dm}}{dN}=-2\tilde{V}_{dm}-\frac{1}{H_{0}\sqrt{\Omega_{r}}}\Psi, \label{peNvdm}
\end{eqnarray}
\end{widetext}
where $\tilde{V}\equiv V/a$. If $\Pi_{de}$, $\Pi_{dm}$ and $w$ are constants in the early epoch, we can obtain analytical solutions of these equations, which in the general case for all three types of the DE-DM interaction are:
\begin{eqnarray}
 & & \delta_{de}=-\frac{3}{2}(1+w+\Pi_{de})\Psi+\delta_{de}^{*}, \label{initdde} \\
 & & \delta_{dm}=-\frac{3}{2}(1-\Pi_{dm})\Psi+\delta_{dm}^{*}, \label{initddm} \\
 & & \delta_{r}=-2\Psi, \label{initdr} \\
 & & \tilde{V}_{de}=-\frac{1}{2H_{0}\sqrt{\Omega_{r}}}\Psi+\tilde{V}_{de}^{*}, \label{initvde} \\
 & & \tilde{V}_{dm}=-\frac{1}{2H_{0}\sqrt{\Omega_{r}}}\Psi+\tilde{V}_{dm}^{*}, \label{initvdm} \\
 & & \tilde{V}_{r}=-\frac{1}{2H_{0}\sqrt{\Omega_{r}}}\Psi, \label{initvr}
\end{eqnarray}
where $\delta_{de}^{*}$, $\delta_{dm}^{*}$, $\tilde{V}_{de}^{*}$, $\tilde{V}_{dm}^{*}$ are the deviations from adiabatic constant solutions. For all the three cases of the DE-DM interaction considered in this work $\delta_{dm}^{*}=0$, but in the general case it could be the non-zero function of $N$. In the non-interacting case, these non-adiabatic perturbations vanish fast. But when there is a DE-DM interaction, then they can increase over time, so the solution (\ref{initdde})-(\ref{initvr}) is not stable at supper-horizon scales in the radiation dominated epoch. As mentioned before, when we have the quintessence dark energy with the constant EoS parameter being close to $-1$, for a DE-DM interaction dependent on the density of dark matter (\ref{fdm}), the super-horizon non-adiabatic mode of dark energy perturbations in the radiation dominated epoch are unstable \cite{Valiviita2008}. In the model of dynamical dark energy which we study, the EoS parameter is variable in time but in the early epoch (both in the quintessence and phantom models) it can be assumed constant. So we can use the initial conditions (\ref{initdde})-(\ref{initvr}) but now the dark energy EoS parameter in the early epoch does not need to be close to $-1$, while staying close to this value in the modern epoch.

To derive the stability conditions for the perturbations of interacting dark energy in the radiation dominated epoch, we use the Liénard-Chipart criterion \cite{Lienard} for the analysis of non-adiabatic solutions of perturbation equations (\ref{peNdde})-(\ref{peNvdm}). It must be noted that these conditions must also be used with the density positivity conditions of the dark components obtained in \cite{Neomenko2016}.

\subsection*{Interactions $\bar{f}=\alpha\bar{\rho}_{cr}$ and $\bar{f}=\beta\bar{\rho}_{de}$}

For the interaction model independent of densities of dark components (\ref{fcr})  perturbations of dark energy are stable in the radiation dominated epoch for $c_{a}^{2}<0$ and $0\leq c_{s}^{2}\leq1$, both for quintessence and phantom dark energy.

For the interaction proportional to the density of dark energy (\ref{fde}) ($0\leq c_{s}^{2}\leq1$) the stability condition for dark energy perturbations in the radiation dominated epoch for $-1<c_{a}^{2}<0$ is
$$\beta<\min\biggl[\frac{1+c_{a}^{2}}{1+c_{s}^{2}}\biggl(\frac{2}{3}-c_{a}^{2}\biggr), \quad \frac{2}{3}(1+c_{a}^{2})\biggr],$$
and for $-\infty <c_{a}^{2}<-1$ the dark energy is stable. This result is in agreement with those obtained in \cite{Jackson2009}.

\subsection*{Interaction $\bar{f}=\gamma\bar{\rho}_{dm}$}

For the interaction proportional to the density of dark matter (\ref{fdm}) the quintessence dark energy is stable if $\gamma>\gamma_{0}$ for $b_{1}<c_{a}^{2}<\min(0, \quad b_{2})$. If $c_{a}^{2}$ is not in that range, then there is an additional condition $\gamma\in(-\infty, \quad \gamma_{1}) \cup (\gamma_{2}, \quad \infty)$, where
\begin{eqnarray}
& & b_{1,2}=-\frac{d_{2}\pm\sqrt{d_{2}^{2}-4d_{1}d_{3}}}{2d_{1}}, \nonumber \\
& & d_{1}=9(1+c_{s}^{2})^{2}+12(1+c_{s}^{2})+4, \nonumber \\
& & d_{2}=12(1+c_{s}^{2})-24c_{s}^{2}(1+c_{s}^{2})+8, \nonumber \\
& & d_{3}=4-24c_{s}^{2}(1+c_{s}^{2}), \nonumber \\
& & \gamma_{0}=-c_{a}^{2}-\frac{1+c_{a}^{2}}{2+c_{s}^{2}+c_{a}^{2}}\biggl(\frac{2}{3}-c_{a}^{2}\biggr), \nonumber \\
& & \gamma_{1,2}=-\frac{1}{2}c_{a}^{2}-\frac{1}{3}\frac{1+c_{a}^{2}}{1+c_{s}^{2}}\mp \nonumber \\
& & \mp\frac{1}{18}\frac{1+c_{a}^{2}}{1+c_{s}^{2}}\sqrt{\biggl(9\frac{1+c_{s}^{2}}{1+c_{a}^{2}}c_{a}^{2}-6\biggr)^{2}-216\frac{1+c_{s}^{2}}{1+c_{a}^{2}}(c_{s}^{2}-c_{a}^{2})}.\nonumber
\end{eqnarray}
For phantom dark energy stability condition is $\gamma<\gamma_{0}$ ($-2-c_{s}^{2}<c_{a}^{2}<-1$) and $\gamma>\gamma_{0}$ ($c_{a}^{2}<-2-c_{s}^{2}$) with additional condition $\gamma\in(-\infty, \quad \gamma_{1}) \cap (\gamma_{2}, \quad \infty)$. This result is also in agreement with those obtained in \cite{Valiviita2008}.

\section{Numerical results}

The system of equations for the density and velocity perturbations of dark energy, dark matter and radiation (\ref{pedde})-(\ref{pepsi}) is integrated using the Fortran subroutine dverk.f \cite{dverk} based on the Runge-Kutta-Verner fifth and sixth order method. The initial value of gravitational potential was taken as $\Psi=-4.25\cdot 10^{-5}$ and the initial scale factor as $a=10^{-10}$.

In Figs. \ref{fig:cr1} - \ref{fig:dm1} the evolution of the Fourier mode $k=0.01Mpc^{-1}$ amplitude of cosmological perturbations is shown for the quintessence and phantom dark energy models with a non-gravitational interaction. For convenience, we present the effective momentums of dark components:
$$\nu_{de}=(1+w)V_{de}+\Pi_{de}V_{T}, \quad \nu_{dm}=V_{dm}-\Pi_{dm}V_{T},$$
instead of velocity perturbations $V_{de}$ and $V_{dm}$ accordingly.

One can see, that in the conformal-Newtonian gauge, the amplitudes of the dark matter density perturbations (bottom panels in the left columns of each figure) at the super-horizon stage ($a<0.001$) are constant, while the amplitudes of the dark matter velocity perturbations increases proportional to $a$ (bottom panels in the right columns of each figure). This is well known from classical papers \cite{Bardeen1980} and \cite{Kodama1984}. The perturbations of quintessential dark energy at this stage evolve similarly and independently on the type of interaction (upper panels of upper part of figures).

\subsection*{Interaction $\bar{f}=\alpha\bar{\rho}_{cr}$}
 
Fig. \ref{fig:cr1} shows the evolution of the density and effective momentum perturbations of dark energy and dark matter for coupling (\ref{fcr}). One can see that the density and effective momentum perturbations of quintessence dark energy $\delta_{de}$ and $\nu_{de}$ change their signs at times closer to the modern epoch for a non-zero interaction parameter. This happens due to the growth of the momentum transfer between dark energy and dark matter which is caused by the non-gravitational interaction. The amplitude of the density perturbations of dark matter increases faster in the matter dominated epoch, due to the impact of the dark coupling on the gravitational potential perturbation, and slower closer to the present epoch due to an increase in the effective pressure of dark matter compared with the non-interacting case. In the phantom dark energy model, compared with the non-interacting case, there are changes in the sign of the density perturbation due to the change of sign of the dark energy effective momentum $\nu_{de}$. The perturbations of dark matter in the phantom dark energy case evolve similarly as in the quintessence case.

\subsection*{Interaction $\bar{f}=\beta\bar{\rho}_{de}$}

Fig. \ref{fig:de1} shows the evolution of the density and effective momentum perturbations of dark energy and dark matter for coupling (\ref{fde}). The oscillations of $\delta_{de}$ and $\nu_{de}$ which arise after entering the Hubble horizon, are due to the impact of the DE-DM interaction on the pressure of dark energy. Their amplitude is proportional to the interaction parameter $\beta$. The dark matter density and velocity perturbations are very sensitive to the value of interaction parameter in the case of quintessence, and practically insensitive in the case of phantom dark energy. This has an obvious explanation: $\Pi_{dm}=\beta\bar{\rho}_{de}/\bar{\rho}_{dm}$ is essentially larger in the past in the first case than in the second one. The greater the value of the interaction parameter, the more quintessential dark energy suppresses the increasing of the perturbation amplitude of dark matter. The initial evolution of the phantom dark energy perturbations strongly increases its dependence on the large values of interaction parameter $\beta$, because at these values $w$ is not constant, so the initial adiabatic conditions are not valid anymore, so we must take the small values of $\beta$. The evolution of $\delta_{de}$ for small values of $\beta$ is defined by the evolution of effective momentum $\nu_{de}$, as in the previous interaction model.

\subsection*{Interaction $\bar{f}=\gamma\bar{\rho}_{dm}$} 

Fig. \ref{fig:dm1} shows the evolution of the density and effective momentum perturbations of dark energy and dark matter for coupling (\ref{fdm}). In this case, the evolution of $\delta_{de}$ and $\nu_{de}$ after entering the Hubble horizon is different for quintessence and phantom: the former oscillates, the latter alters the sign of amplitude. Also we see a significant difference in the behaviour of the non-interacting and interacting phantom dark energy models in the early epoch. This is due the fact that for this
type of coupling, the phantom dark energy at the beginning behaves like quintessence, its density decreases, and only after some time it begin to increase (see Fig. 5b in \cite{Neomenko2016}). The smaller is the interaction parameter $\gamma$ the shorter is the period of the decrease of the phantom dark energy density. But models of dark energy similarly affect the amplitude of the density and velocity perturbations of dark matter: the faster increase just after entering the Hubble horizon and the slower increase at later stages compared with minimally coupled models of dark energy. This behaviour of dark matter perturbations follows from Eq. (\ref{peddm}) and the fact that for this type of interaction $\Pi_{dm}=\gamma$ is for both models of dark energy and any $a$.

\section*{CONCLUSIONS}

We studied the evolution of cosmological perturbations in the model with dynamical dark energy coupled with dark matter by an additional non-gravitation interaction. This model, with variable EoS parameter, behaves in such way that in the early epoch EoS can be considered as constant. The dark energy perturbations here are free from non-adiabatic instabilities at super-horizon scales in the radiation dominated epoch. The stability conditions for the perturbations of dark energy are in the form of constraints on the value of the interaction parameter obtained in an explicit form. We studied three cases of dark coupling: independent on the densities of dark components, proportional to the density of dark energy and proportional to the density of dark matter. In all cases the impact of the interaction on the evolution of dark energy perturbations is larger than on the dark matter ones. The impact of the interaction on the dark matter cosmological perturbations is similar for all cases: after the perturbations enter the Hubble horizon, their growth is faster, and closer to present epoch it is slower. The strength of the impact depends on the type of interaction, on the value of its parameter ($\alpha$, $\beta$, $\gamma$) and the type of the dark energy model. The results can be useful for establishing of the observational constraints on the nature of dark components and possible interaction between them.

\section*{ACKNOWLEDGEMENTS}

This work was supported by the project of Ministry of Education and Science of Ukraine ``Formation and characteristics of elements of the structure of the multicomponent Universe, gamma radiation of supernova remnants and observations of variable stars'' (state registration number 0119U001544).


\begin{thebibliography}{0}
\expandafter\ifx\csname natexlab\endcsname\relax\def\natexlab#1{#1}\fi
\expandafter\ifx\csname bibnamefont\endcsname\relax
  \def\bibnamefont#1{#1}\fi
\expandafter\ifx\csname bibfnamefont\endcsname\relax
  \def\bibfnamefont#1{#1}\fi
\expandafter\ifx\csname citenamefont\endcsname\relax
  \def\citenamefont#1{#1}\fi
\expandafter\ifx\csname url\endcsname\relax
  \def\url#1{\texttt{#1}}\fi
\expandafter\ifx\csname urlprefix\endcsname\relax\def\urlprefix{URL }\fi
\providecommand{\bibinfo}[2]{#2}
\providecommand{\eprint}[2][]{\url{#2}}

\end{thebibliography}


\begin{thebibliography}{}

\bibitem{Riess1998} A. G. Riess et al, The Astronomical Journal 116, 3 (1998);

\bibitem{Perlmutter1999} S. Perlmutter et al., The Astrophysical Journal, 517, 2, 565-586 (1999);

\bibitem{Colin2019} J. Colin, R. Mohayaee, M. Rameez, S, Sarkar, A\&A 631, L13 (2019);

\bibitem{Kang2020} Y. Kang, Y.-W. Lee, Y.-L. Kim, C. Chung, C. H. Ree, ApJ 889, 1 (2020);

\bibitem{DiValentino2020} E. Di Valentino, A. Melchiorri, J. Silk, Nature Astronomy 4, 196-203 (2020);

\bibitem{Nadathur2020} S. Nadathur, W. J. Percival, F. Beutler, H. A. Winther, arXiv:2001.11044;

\bibitem{Amendola2000} L. Amendola, Phys. Rev. D. 62, 043511, (2000);

\bibitem{Zimdahl2001} W. Zimdahl, D. Pavon, L. P. Chimento, Phys. Lett. B. 521, 3-4, 133-138 (2001);

\bibitem{Amendola2007} L. Amendola, G. C. Campos, R. Rosenfeld, Phys. Rev. D. 75, 083506 (2007);

\bibitem{Bolotin2013}  Yu. L. Bolotin, A. Kostenko, O. A. Lemets, D. A. Yerokhin, Int. J. Mod. Phys. D. 24, 1530007 (2015);

\bibitem{Chimento2010} L. P. Chimento, Phys. Rev. D 81, 043525 (2010);

\bibitem{DiValentino2017} E. Di Valentino, A. Melchiorri, O. Mena, Phys. Rev. D. 96, 043503 (2017);

\bibitem{Rui2018} R. An, C. Feng, B. Wang, JCAP 02, 038 (2018);

\bibitem{DiValentino2019} E. Di Valentino, A. Melchiorri, O. Mena, and S. Vagnozzi, arXiv:1908.04281;

\bibitem{Valiviita2008} J. Väliviita, E. Majerotto, R. Maartens, JCAP 07, 020 (2008);

\bibitem{Jackson2009} B. M. Jackson, A. Taylor, A. Berera, Phys. Rev. D 79, 043526 (2009)

\bibitem{Majerotto2010} E. Majerotto, J. Väliviita, R. Maartens, MNRAS 402, 4, 2344-2354 (2010);

\bibitem{Sharov2017} G. S. Sharov, S. Bhattacharya, S. Pan, R. C. Nunes, S. Chakraborty, MNRAS 466, 3 (2017);

\bibitem{WYang2018} W. Yang, A. Mukherjee, E. Di Valentino, S. Pan, Phys. Rev. D 98, 123527 (2018);

\bibitem{Yang2019a} W. Yang, N. Banerjee, A. Paliathanasis, S. Pan, Physics of the Dark Universe 26, 100383 (2019);

\bibitem{Bonici2019} M. Bonici, N. Maggiore, Eur. Phys. J. C 79, 672 (2019);

\bibitem{Goswami2019} G. K. Goswami, A. Pradhan, A. Beesham, Pramana 93, 89 (2019);

\bibitem{Novosyadlyj2010} B. Novosyadlyj, O. Sergijenko, S. Apunevych, V. Pelykh, Phys. Rev. D. 82, 103008 (2010);

\bibitem{Novosyadlyj2012} B. Novosyadlyj, O. Sergijenko, R. Durrer, V. Pelykh, Phys. Rev. D. 86, 083008 (2012);

\bibitem{Sergijenko2015} O. Sergijenko, B. Novosyadlyj, Phys. Rev. D. 91, 083007 (2015);

\bibitem{Neomenko2016} R. Neomenko, B. Novosyadlyj, Kinemat. Phys. Celest. Bodies 32, 157 (2016);

\bibitem{SWang2008} S. Wang, Y. Zhang, Phys. Lett. B 669, 3-4, 201-205 (2008);

\bibitem{Caldera2009} G. Caldera-Cabral, R. Maartens, L. A. Ureña-López, Phys. Rev. D 79, 063518 (2009);

\bibitem{Feng2020} L. Feng, H. Li, J. Zhang, X. Zhang, Sci. China Phys. Mech. Astron. 63, 220401 (2020);

\bibitem{Zhou2009} J. Zhou, B. Wang, D. Pavón, E. Abdalla, Mod. Phys. Lett. A 24, 21, 1689-1698 (2009);

\bibitem{delCampo2006} S. del Campo, R. Herrera, G. Olivares, D. Pavón, Phys. Rev. D 74, 023501 (2006);

\bibitem{delCampo2009} S. del Campo, R. Herrera, D. Pavón, JCAP 01, 020 (2009);

\bibitem{Feng2019} L. Feng, J. Zhang, X. Zhang, Physics of the Dark Universe 23, 100261 (2019)

\bibitem{Wang2007} B. Wang, J. Zang, C. Lin, E. Abdalla, S. Micheletti, Nuclear Physics B 778, 1-2, 69-84 (2007);

\bibitem{Rosenfeld2007} R. Rosenfeld, Phys. Rev. D 75, 083509 (2007);

\bibitem{Gavela2010} M. B. Gavela, L. Lopez Honorez, O. Mena, S. Rigolin, JCAP 11, 044 (2010);

\bibitem{Neomenko2017} R. Neomenko, B. Novosyadlyj, O. Sergijenko, J. Phys. Stud. 21, 3 (2017);

\bibitem{NTK} B. Novosyadlyj, M. Tsizh, Yu. Kulinich, Gen. Rel. Grav. 48, 30 (2016);

\bibitem{Lienard} A. Liénard, H. Chipart, J. Math. Pures Appl. 10, 291–346 (1914);

\bibitem{dverk} http://www.cs.toronto.edu/NA/dverk.f.gz

\bibitem{Bardeen1980} J. M. Bardeen, Phys. Rev. D 22, 1882 (1980);

\bibitem{Kodama1984} H. Kodama, M. Sasaki, Progress of Theoretical Physics Supplement 78, 1-166 (1984);

\end{thebibliography}
\end{document}